\documentclass[prc,preprint,showpacs,preprintnumbers,amsmath,amssymb]{revtex4}

\usepackage[mathscr]{eucal}
\usepackage{graphicx}
\def\slr#1{\setbox0=\hbox{$#1$}           
   \dimen0=\wd0                                 
   \setbox1=\hbox{/} \dimen1=\wd1               
   \ifdim\dimen0>\dimen1                        
      \rlap{\hbox to \dimen0{\hfil/\hfil}}      
      #1                                        
   \else                                        
      \rlap{\hbox to \dimen1{\hfil$#1$\hfil}}   
      /                                         
   \fi}

\def\myvecint#1{\!\int\!\!\frac{d{#1}}{(2\pi)^3}\,}
\def\gev#1{ GeV${}^{#1}$}
\def\be{\begin{eqnarray}}
\def\ee{\end{eqnarray}}

\def\etal{\emph{et al.}}

\renewcommand{\theequation}%
    {\arabic{section}.\arabic{equation}}
\makeatletter \@addtoreset{equation}{section} \makeatother

\begin{document}

\preprint{BCCNT: 05/314/333}

\title{Relativistic Calculation of the Width of the $\Theta^+(1540)$ Pentaquark}

\author{Hu Li}
\author{C. M. Shakin}
\email[email:]{cshakin@brooklyn.cuny.edu}

\affiliation{%
Department of Physics\\
Brooklyn College of the City University of New York\\
Brooklyn, New York 11210
}%

\author{Xiangdong Li}
\affiliation{%
Department of Computer System Technology\\
New York City College of Technology of the City University of New
York\\
Brooklyn, New York 11201 }%

\date{\today}

\begin{abstract}
We calculate the width of the $\Theta^+(1540)$ pentaquark in a
relativistic model in which the pentaquark is considered to be
composed of a scalar diquark and a spin 1/2 triquark. We consider
both positive and negative parity for the pentaquark. There is a
single parameter in our model which we vary and which describes
the size of the pentaquark. If the pentaquark size is somewhat
smaller than that of the nucleon, we find quite small widths for
the pentaquark of about 1 MeV or less. Our model of confinement
plays an important role in our analysis and makes it possible to
use Feynman diagrams to describe the decay of the pentaquark.
\end{abstract}


\maketitle

\section{INTRODUCTION}

There has been a great deal of interest in the recent observation of
a quite narrow resonance which decays to a nucleon and a kaon
\cite{Na03,Ba03a,Al04,Ak04,Ku04,St03,Ba03b,Ab04,ch03}. These states
have been interpreted as pentaquarks which are likely to have a
$udud\bar s$ structure \cite{Ja03}. (The extremely narrow width of
the state, the $\Theta^+(1540)$, has led some to question whether
the state exists in nature.) In the present work we will describe
the calculation of the width of the $\Theta^+(1540)$ in a
relativistic quark model which includes a model of confinement that
we have introduced previously in our study of meson and nucleon
structure.

\begin{figure}
\includegraphics[bb=0 355 519 623, angle=0, scale=0.7]{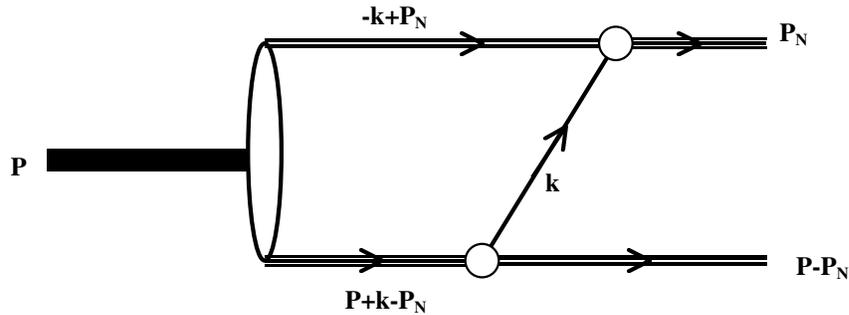}%
\caption{\label{f333.1}In this figure the heavy line denotes the
pentaquark and the line of momentum $-k+P_N$ denotes an
on-mass-shell diquark. The line of momentum $k$ represents the quark
and $P+k-P_N$ is the momentum of the triquark. In the final state we
have a nucleon of momentum $P_N$ and a kaon of momentum $P-P_N$.}
\end{figure}

A number of reviews have appeared. In Ref. \cite{Ca04} the
experimental evidence for the pentaquark is reviewed. Theoretical
and experimental developments are reviewed in Refs.
\cite{Zh04,St04,Cl03,Na04a}. A number of theoretical papers have
also appeared \cite{Ja03,Sh04,Se05,Na04b,Ma04,Ka04,Ka03}. We are
particularly interested in the work of Ref. \cite{Ka03} in which a
diquark-triquark model is introduced to describe the pentaquark. We
will make use of a variant of that model in the present work, since
that model lends itself to the analysis of the pentaquark decay
using Feynman diagrams. The diagram we consider is presented in Fig.
\ref{f333.1}. There the pentaquark is represented by the heavy line
with momentum $P$. The pentaquark is composed of a diquark of
momentum $-k+P_N$ and a triquark of momentum $P+k-P_N$. The triquark
emits a quark ($u$ or $d$) which combines with the diquark to form a
nucleon of momentum $P_N$. The final-state kaon of momentum $P-P_N$
is emitted along with the quark at the triquark vertex.

We have studied the structure of the nucleon in a quark-diquark
model in Ref. \cite{Sz96}. In that work we considered both scalar
and axialvector diquarks, however, in this work we will limit our
considerations to a nucleon composed of a quark and a scalar
diquark. The nucleon vertex is described in Ref. \cite{Sz96} for the
case in which the diquark is placed on mass shell. (We remark that
in Fig. \ref{f333.1} only the quark of momentum $k$ and the triquark
of momentum $P+k-P_N$ will be off mass shell in our analysis.) The
mass of the scalar diquark was taken to be 400 MeV in Ref.
\cite{Sz96}. We use that value here and also take the triquark to
have a mass of 800 MeV. Therefore, it is clear that we need to
introduce a model of confinement for the pentaquark which has a mass
of 1540 MeV. Our (covariant) confinement model which we have used
extensively in other works \cite{Ce99a,Ce99b,Ce99c} will be
discussed in the next section.

The organization of our work is as follows. In Section II we will
describe our model of confinement and in Section III we will
calculate the width of the pentaquark, assuming it has negative
parity. In Section IV we consider the width in the case the
pentaquark has positive parity. Finally, Section V contains further
discussion and conclusions.

\section{A Model of Confinement}

As stated in the last section, our pentaquark is composed of a
scalar diquark of mass 400 MeV and a triquark of mass 800 MeV. Since
the mass of the pentaquark under consideration is 1540 MeV, the
pentaquark would decay into its constituents in the absence of a
model of confinement. Similarly, the nucleon of mass 939 MeV could
decay into the scalar diquark of mass 400 MeV and a quark whose mass
we take to be 350 MeV in this work.

In earlier work we have introduced a confining interaction which
served to prevent the decay of mesons or nucleons into their
constituents. Our covariant confinement model is described in a
series of our papers \cite{Ce99a,Ce99b,Ce99c}. In that model we
solve a linear equation for a confining vertex function, $\Gamma$.
This function has the following property. Consider the decay
$A\rightarrow B+C$, in which the hadrons $A$ and $C$ are on mass
shell. If we include the confining vertex we find the amplitude
has a zero when particle $B$ goes on mass shell, so that the
amplitude for $A$ to decay into two on-mass-shell particles ($B$
and $C$) is zero.

For the decay $A\rightarrow B+C$ we may introduce a wave function
that may be expressed in terms of the momentum of the off-shell
particle $B$. If $B$ is a scalar, we have \be\label{e333.2.1}
\Psi_B(k)=\frac1{k^2-m_B^2}\Gamma_B(k)\,.\ee Note that the ratio
of $\Gamma_B(k)$ to $(k^2-m_B^2)$ is an ordinary function which
may often be well represented by a Gaussian function. (It is not
necessary to include an $i\epsilon$ in the denominator of Eq.
(\ref{e333.2.1}).)

In the case of the nucleon we may consider the decay into a quark
and a diquark. In Ref. \cite{Sz96} we considered both scalar and
axialvector diquarks, but for simplicity we will limit ourselves
to only the scalar diquark. The relevant wave function in this
case was given in Eq. (3.3) of Ref. \cite{Sz96}:
\be\Psi_S(P,k,s,t)=\widetilde\Psi_{(1)}(P,k)\frac{2m_q\Lambda^{(+)}(\vec
k)}{\sqrt{2E(\vec k)\left(E(\vec k)+m_q\right)}}u_N(P,s)\chi_t\ee
which we will simplify for the present work to read
\be\label{e333.2.3}\Psi_N(P_N,k,s,t)=\widetilde\Psi_N(P_N,k)\Lambda^{(+)}(\vec
k)u_N(P_N,s)\chi_t\,.\ee The function $\widetilde\Psi_N(P,k)$ is
represented in Fig. 5 of Ref. \cite{Sz96} by a dashed line. That
function is well approximated by a Gaussian function which we will
record at a latter point in our discussion. The factor of
$\Lambda^{(+)}(\vec k)=(\slr k_{on}+m_q)/2m_q$ arises from an
approximation made in Ref. \cite{Sz96}. (Here $k_{on}=(E(\vec
k),\,\vec k)$ with $E(\vec k)=[\vec k^{\,2}+m_q^2]^{1/2}$.) In
Ref. \cite{Sz96} the quark propagator was written as \be
-iS(k)=\frac{m_q}{E_q(k)}\left[\frac{\Lambda^{(+)}(\vec
k)}{k^0-E_q(\vec k)}-\frac{\Lambda^{(-)}(-\vec k)}{k^0+E_q(\vec
k)}\right]\,.\ee The second term was neglected in our formalism
when we studied the nucleon. (Thus we limited our analysis to
positive-energy quark spinors.) Since we wish to make use of the
nucleon wave function determined in Ref. \cite{Sz96}, we will
continue to include the projection $\Lambda^{(+)}(\vec k)$ in our
formalism. Here $k$ is the quark momentum.

As stated earlier, the diquark of momentum $-k+P_N$ in Fig.
\ref{f333.1} will be placed on mass shell, so that $k^0=E_N(\vec
P_N)-E_D(\vec P_N-\vec k)$, where $E_D(\vec P_N-\vec k)=[(\vec
P_N-\vec k)^{\,2}+m_D^2]^{1/2}$ and $E_N(\vec P_N)=[\vec
P_N^{\,2}+m_N^2]^{1/2}$. That approximation is achieved by writing
\be \frac1{(P_N-k)^2-m_D^2+i\epsilon}\longrightarrow -2\pi
i\delta^{(+)}[(P_N-k)^2-m_D^2]\,,\ee as described in detail in
Ref. \cite{Ce86}. Note that \be
\delta^{(+)}[(P_N-k)^2-m_D^2]=\frac1{2E_D(\vec P_N-\vec
k)}\,\delta[P_N^0-k^0-E_D(\vec P_N-\vec k)]\,.\ee The
on-mass-shell specification used here arises when performing an
integral in the complex $k^0$ plane \cite{Ce86}.

\section{Calculation of the Width of a Negative Parity Pentaquark}

We consider the diagram shown in Fig. \ref{f333.1}. Recall that the
heavy line of momentum $P$ denotes the pentaquark. The line carrying
momentum $-k+P_N$ is the on-mass-shell scalar diquark and the line
with momentum $P+k-P_N$ is the triquark. The momentum $k$ is that of
an up or down quark, $P-P_N$ is the momentum of the nucleon. The
pentaquark, nucleon and kaon are on mass shell. In our analysis the
diquark is also on mass shell, so that only the triquark and quark
propagate off mass shell, as noted earlier. (As stated in the last
section, the on-shell characterization of the diquark arises when we
complete the $k^0$ integral in the complex $k^0$ plane.)

We now make use of the formula \cite{Bj64} \be
d\Gamma=|\mathfrak{M}|^2\frac{d^3k_1}{(2\pi)^3}\,\frac{m_N}{E_N(\vec
k_1)}\,\frac{d^3k_2}{(2\pi)^3}\,\frac1{2E_K(\vec k_2)}\,
(2\pi)^4\delta^4(P-k_1+k_2)\ee where $\vec k_1$ and $\vec k_2$ are
the momenta of the outgoing particles. We may put $\vec k_1=\vec
P_N$ and $\vec k_2=\vec P-\vec P_N=-\vec P_N$ for a pentaquark at
rest. (It is convenient to take $\vec P_N$ along the $z$ axis when
calculating the width.)

In writing our expression for $\Gamma$ we will represent the product
of the quark propagator and nucleon vertex function by the nucleon
wave function of Eq. (\ref{e333.2.3}). In a similar fashion we will
represent the product of the triquark propagator and the
pentaquark-triquark vertex function by a triquark wave function. (In
the vertex, the diquark is on mass shell.) We then have to specify
the vertex function of the triquark which describes the decay into
the quark of momentum $k$ and the kaon. That scalar part of the
vertex is usefully written as \be
\Gamma_T(k)=\Psi_T(k)(k^2-m_q^2)\,.\ee Thus, the wave functions
$\Psi_N(k)$, $\Psi_T(k)$ and $\Psi_\Theta(P_N-k)$ will appear in our
expression for $\Gamma_\Theta$. Note that \be \vec
P_N^2=\left(\frac{m_\Theta^2-m_N^2+m_K^2}{2m_\Theta}\right)-m_K^2\ee
which yields $\vec P_N^2=0.0396$ \gev2, or $|\vec P_N|=0.199$ GeV.

We find that the width is given by
\be\label{e333.3.4}\Gamma_\Theta&=&\frac12\myvecint {\vec
k}\frac1{2E_D(\vec k-\vec P_N)}\myvecint{\vec
k^{\,\prime}}\frac1{2E_D(\vec k^{\,\prime}-\vec
P_N)}\frac1{N_\Theta N_T N_N}\frac1{(2m_N)(2m_\Theta)}\\\nonumber
&\times&\Psi_N(\vec k)\,\Psi_N(\vec
k^{\,\prime})\,\Psi_\Theta(\vec P_N-\vec k)\,\Psi_\Theta(\vec
P_N-\vec k^{\,\prime})\,(k^2-m_q^2)\Psi_T(\vec
k)\,(k^{\prime2}-m_q^2)\Psi_T(\vec k^{\,\prime})\\\nonumber
&\times&\mbox{Tr}\left[\left(\slr k-\slr P_N+\slr
P+m_T\right)\left(\slr k_{on}+m_q\right)\left(\slr
P_N+m_N\right)\gamma^0\right.\\\nonumber &&\left. \left(\slr
k_{on}^\prime+m_q\right)\left(\slr k^\prime-\slr P_N+\slr
P+m_T\right)\gamma^0\left(\slr P+m_\Theta\right)\right]\rho\,,\ee
where $\rho$ is the phase space factor. We find \be
\rho=\frac1{2\pi}\frac{m_N}{m_\Theta}|\vec P_N|\\=0.193
\,\;\mbox{GeV}\,,\ee since $|\vec P_N|\simeq0.199$ GeV.

The factor $\slr P_N+m_N$ and $\slr P+m_\Theta$ in the trace arise
from the relations \be\frac{\slr P_N+m_N}{2m_N}=\sum_{s_N}u_N(\vec
P_N,s_N)\bar u_N(\vec P_N,s_N)\ee and \be\frac{\slr
P+m_\Theta}{2m_\Theta}=\sum_s u_\Theta(\vec P,s)\bar u_\Theta(\vec
P,s)\,.\ee The factors $(1/N_\Theta)^{1/2}$, $(1/N_T)^{1/2}$ and
$(1/N_N)^{1/2}$ serve to normalize the wave functions. We
determine that $N_N=0.316$, $N_T=0.0673$, and calculate $N_\Theta$
for each choice of the pentaquark wave function. We write, with
$k=|\vec k|$, \be\Psi_N(\vec k)=\frac1{\sqrt{N_N}}\left(y_0+\frac
A{w\sqrt{\dfrac\pi2}}\,\,e^{-\dfrac{2(k-k_c)^2}{w^2}}\right)\ee
and \be\Psi_T(\vec k)=\frac1{\sqrt{N_T}}\left(y_0+\frac
A{w\sqrt{\dfrac\pi2}}\,\,e^{-\dfrac{2(k-k_c)^2}{w^2}}\right)\,.\ee
We have determined $y_0$, $A$, $k_c$ and $w$ from a fit to the
wave function given in Fig. 5 of Ref. \cite{Sz96}. We find
$y_0=-3.66$, $k_c=-0.013$ GeV, $w=0.660$ GeV and $A=27.73$ GeV.
For the pentaquark we write \be\Psi_\Theta(\vec k-\vec
P_N)=\frac1{\sqrt{N_\Theta}}\frac
A{w_\Theta\sqrt{\dfrac\pi2}}\,\,e^{-\dfrac{2(\vec k-\vec
P_N)^2}{w_\Theta^2}}\,,\ee where $w_\Theta$ is a variable in our
analysis. (Note that $N_\Theta$ depends upon the choice of
$w_\Theta$.) Here $\vec k-\vec P_N$ is the relative momentum of
the diquark and triquark when $\vec P=0$.

\begin{figure}
\includegraphics[bb=0 0 280 235, angle=0, scale=1]{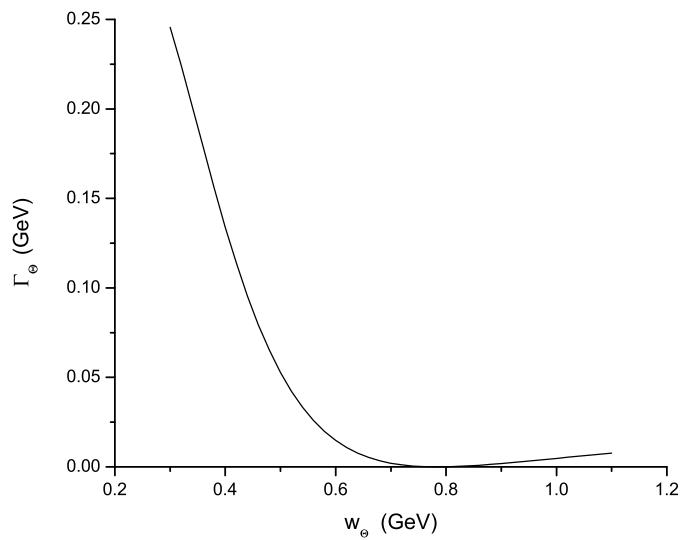}%
\caption{\label{f333.2}The figure shows the width of the
pentaquark as a function of the parameter, $w_\Theta$, which
governs the extent of the pentaquark wave function in momentum
space. In the case of the nucleon we found $w=0.66$ GeV for the
wave function calculated in Ref. \cite{Sz96}. In this calculation
the pentaquark had negative parity as did the triquark.}
\end{figure}

In Fig. \ref{f333.2} we present the results of our calculation of
$\Gamma_\Theta$ as a function of $w_\Theta$. Quite small values
are obtained in the region 0.7 GeV $<w_\Theta<$ 0.9 GeV. The
region 0.6 GeV to 1 GeV is shown in Fig. \ref{f333.3} where it may
be seen that the width has a maximum of about 15 MeV. The minimum
values are very small and it is quite difficult to make an
accurate calculation of widths significantly less than 1 MeV. (We
have calculated five-dimensional integrals with 40 points for each
variable. Thus the number of points calculated is
$(40)^5\sim10^8$.)

\section{Calculation of the Width of a Positive Parity Pentaquark}

In the case of a positive parity pentaquark we assume that the
pentaquark decays to a positive parity diquark and a positive
parity triquark. As in Section III, the triquark and diquark have
zero relative angular momentum. In this case we need to insert
factors of $i\gamma_5$ at the triquark-kaon vertex where the quark
of momentum $k$ is emitted. In addition the calculation of the
normalization factor, $N_T$, is modified. In Eq. (\ref{e333.3.4}),
the trace becomes \be \mbox{Trace}&=&\mbox{Tr}\left[(\slr k-\slr
P_N+\slr P+m_T)i\gamma_5(\slr k_{on}+m_q)(\slr
P_N+m_N)\gamma^0\right.\\\nonumber &&\left. (\slr
k_{on}^\prime+m_q)i\gamma_5(\slr k^\prime-\slr P_N+\slr
P+m_T)\gamma^0(\slr P+m_\Theta)\right]\,.\ee We may define
$\widetilde{ \slr k}=\gamma^0\slr k\gamma^0$, etc. Thus \be
\mbox{Trace}&=&\mbox{Tr}\left[(\slr k-\slr P_N+\slr
P+m_T)\gamma_5(\slr k_{on}+m_q)(\slr P_N+m_N)(\widetilde{\slr
k}_{on}^{\,\prime}+m_q)\right.\\\nonumber &&\left.
\gamma_5(\widetilde{\slr k}^{\,\prime}-\widetilde{\slr
P}_N+\widetilde{\slr P}+m_T)(\slr
P+m_\Theta)\right] \\
&=& \mbox{Tr}\left[(\slr k-\slr P_N+\slr P+m_T)(-\slr
k_{on}+m_q)(-\slr P_N+m_N)(-\widetilde{\slr
k}_{on}^{\,\prime}+m_q)\right.\\\nonumber &&\left.
(\widetilde{\slr k}^{\,\prime}-\widetilde{\slr
P}_N+\widetilde{\slr P}+m_T)(\slr P+m_\Theta)\right]\,.\ee

In this case we find $N_N=0.316$ and $N_T=0.0201$, where only the
second value has changed relative to the values given in Section
III. In Fig. \ref{f333.4} we show the values obtained for the
width of the positive parity pentaquark, and in Fig. \ref{f333.5}
we show the results of our calculation using a different scale.

\section{discussion}

It is of interest to note that, despite extensive efforts, no
signal of a 1540 MeV pentaquark has been observed in lattice QCD
studies \cite{La05,Cs05}. (These work contain more comprehensive
references to the literature than given here.) In Ref. \cite{Cs05}
pion masses in the range 400-630 MeV are used. The authors state
that they cannot rule out the existence of a pentaquark at the
physical quark mass of $m_\pi=135$ MeV. They also suggest the
possibility is that the pentaquark has a more exotic wave function
than that considered in Ref. \cite{Cs05}.

\begin{figure}
\includegraphics[bb=0 0 280 235, angle=0, scale=1]{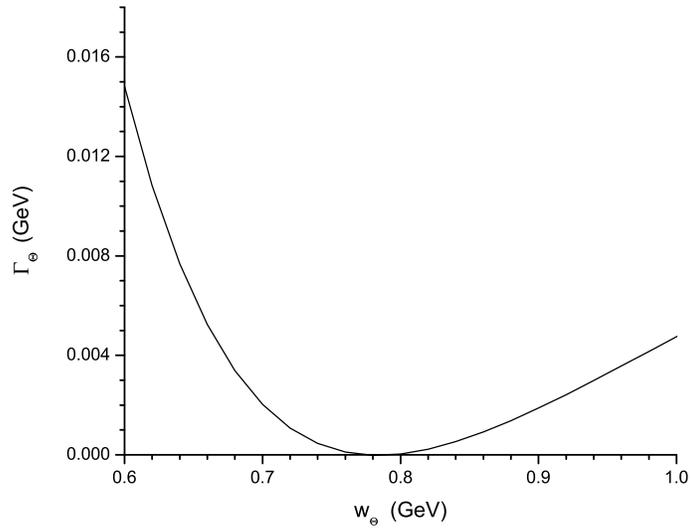}%
\caption{\label{f333.3}The width of the negative parity pentaquark
is shown as a function of the parameter $w_\Theta$. (See Fig.
\ref{f333.2}.) (The quite small values in the vicinity of the
minimum have rather large uncertainties because of the limitation
of the number of points used in our five-dimensional integral
which determines the width.)}
\end{figure}

\begin{figure}
\includegraphics[bb=0 0 280 235, angle=0, scale=1]{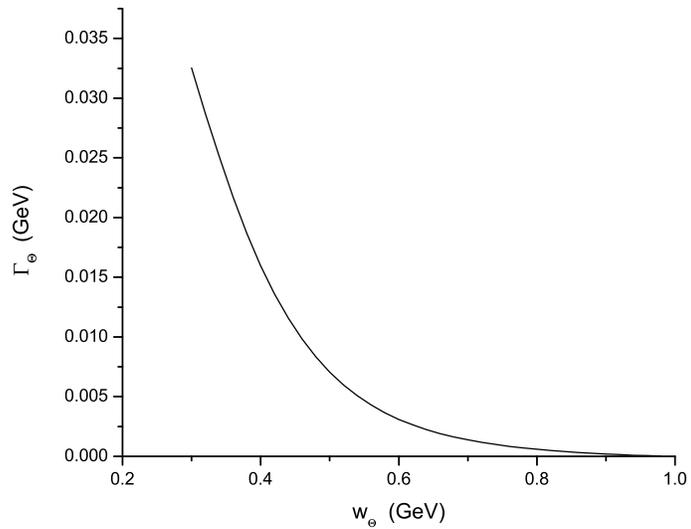}%
\caption{\label{f333.4}The width of a pseudoscalar pentaquark
calculated in our model is shown.}
\end{figure}

\begin{figure}
\includegraphics[bb=0 0 280 235, angle=0, scale=1]{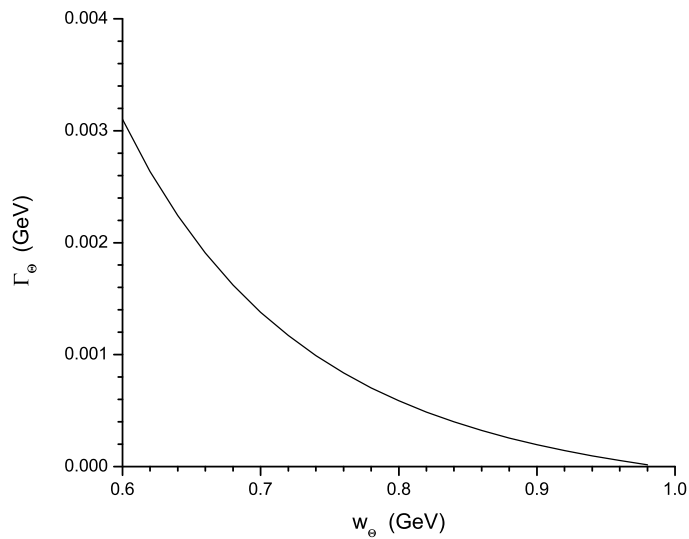}%
\caption{\label{f333.5}The results of the calculation reported in
Fig. 4 are shown using a different scale. Widths of the pentaquark
less than 1 MeV are found for values of $w_\Theta$ greater than
0.72 GeV. (We recall that $w$ for the nucleon was 0.66 GeV.)}
\end{figure}

One feature that has led some researches to question the existence
of the pentaquark is its very small width \cite{Cl03}. We have
shown in the present work that very small widths are obtained in
our model if the pentaquark wave function parameter, $w_\Theta$,
is somewhat larger than that we have found for the nucleon
($w_N=0.66$ GeV) in Ref. \cite{Sz96}. [See Fig. \ref{f333.3}.]



\end{document}